# A search for super-large structures in deep galaxy surveys

© N.V. Nabokov [1,2], Yu. V. Baryshev [1,3]

[1] The Sobolev Astronomical Institute of the St.-Petersburg University
[2] Email: NabokovNikita@yandex.ru
[3] Email: yuba@astro.spbu.ru

**Abstract:** Recent extensive, multi-color deep surveys of galaxies open a possibility to get observational estimation of sizes for the largest structures in the Universe. Photometric redshift accuracy (about $0.03(1+z)$) allows directly study clustering at scales about 1000 Mpc. Thanks to large number of galaxies in each redshift bin one may detect super-large structures if they really exist. Here we show that the observed behavior of the redshift distribution of galaxies in deep surveys such as HUDF and FDF is consistent with existence of super-large structures of luminous matter with scales about 2000 Mpc. We detect a large underdense region in radial galaxy distribution at redshift interval $z = 1.2 \div 2.2$ which separate our "Local Hubble Volume" from the neighboring overdensity region at $z = 2.2 \div 3.5$. This result can also explain the observed deficiency of gamma ray sources at redshift about 2. Observational test on the reality of the supper-large structures may be obtained by organizing sky covering net (cells about 10n x 10n degrees) of very deep narrow angle (1n x 1n arc-minutes) multi-band photometric surveys of galaxies which is achievable for large ground-based telescopes.

## 1. Introduction

An important goal of cosmology is to set an observational limit on the sizes of the largest structures in visible galaxy distribution. Recent deep multi-band photometric surveys of galaxies, such as COMBO-17 [23], COSMOS [13], FDF [10], HUDF [2], deliver a new possibility to estimate a homogeneity scale after which the luminous matter distribution becomes uniform. This is because modern deep surveys contain $10^3 - 10^5$ galaxies with measured multi-band magnitudes, which allow construct radial distribution of galaxies based on photometric redshifts.

The accuracy of measurements of $z_{phot}$ (about 0.06 at $z = 1$) allows to study spatial scales larger than $180 Mpc/h_{100}$ at redshift interval corresponding to the depth of a survey. Intriguingly such scales are much larger than usually accepted "homogeneity scale" or "correlation length" $r_0 = 5 Mpc/h_{100}$ derived from 2-point correlation function analysis in the low redshift galaxy surveys (e.g. for 2dF GRS [9] and for SDSS MG [24]). It is natural to expect a smooth galaxy distribution along z-coordinate for bins larger than $dz = 0.1$, which corresponds to $dr = 300 Mpc/h_{100}$, i.e. 60 times more than "correlation length". The galaxy number fluctuations in bins for observed radial distributions should be restricted by relatively small values for the case of small "correlation length" of the 2-point correlation function [21].

However in all extensive deep surveys (such as COMBO-17, COSMOS, FDF, HUDF), the numbers of galaxies in redshift bins varies essentially, so different authors claimed that they see: "unusual and not representative of the cosmic average" [23], "conspicuous maxima of the galaxy densities" [1], "the effect of cosmic variance are quite apparent" [12]. Such actually observed "large cosmic variance", like Sloan Great Wall with size about $500 Mpc$ [8], open new possibility in observational cosmology related to the direct measurement of galaxy distribution at super-large scales.

Thanks to large number of galaxies ($\Delta N$) in each redshift bin ($\Delta z$), the Poisson's shot noise ($1/\sqrt{\Delta N}$) is less than cosmic variance due to correlated structures in galaxy distribution and it is possible to detect a super-large structures if they really exist. Here we show that the observed behavior of the redshift distribution of galaxies in such deep surveys as HUDF and FDF consistent with existence of super-large structures of luminous matter at scales up to $1000 Mpc$. To exclude possible distortions by hidden selection effects we suggest observational tests which can test the reality of super-large structures in the Universe.

## 2. Photometric redshifts

The main advantage of the photometric redshift is that estimation of distances to large number of galaxies up to very faint magnitude becomes achievable. This gives unique possibility to study super-large structures in galaxy distribution by using direct observations.

The accuracy of photometric redshift estimation depends on the accuracy of photometry, number of observed bands, number of templates for spectral energy distribution (SED) of galaxies, and also on individual features of actually observed galaxies.

The error in redshift estimation may be approximated by the simple relation:
$$\delta z = 0.03(1+z) \tag{1}$$
Hence, at redshift 1 the error is $\delta z = 0.06$. It means that for studies of super-large scales with extension of $z$ about $0.6 \div 1$, the error in photo-z does not much influence the results.

The metric distance in the standard LCDM is given by relation:
$$r(z) = \frac{c}{H_0} \int_{1/1+z}^{1} \frac{dy}{y\sqrt{(\Omega_m^0/y + \Omega_{vac}^0/y^2)}}, \tag{2}$$

where $c$ is the velocity of light, and we use for $h = h_{100} = 0.72 = H_0/100 km/s/Mpc$, $H_0 = 72 km/s/Mpc$, $\Omega_m = 0.3, \Omega_{vac} = 0.7$.

The linear scales which may be directly studied with the above accuracy of photometric redshifts are restricted only by the depth of a deep survey. An example of linear extension of redshift intervals centered at $z = 1$ is presented in Table 1.

| $\Delta z$ | 0.1 | 0.2 | 0.3 | 0.5 | 1.0 |
|---|---|---|---|---|---|
| $\Delta r (Mpc)$ | 237 | 474 | 711 | 1188 | 3114 |

Table 1. The linear sizes $\Delta r$ of scales, which correspond to redshift intervals $\Delta z$ centered at $z = 1$ for the standard parameters of LCDM.

An important strong point of the photometric redshift method is that it is based on continuum energy distribution of galaxies and therefore does not depend on visibility of spectral lines in galaxy spectra at all studied redshifts. However it depends on the deflection of real SED from available templates, and also on sensitivity and common calibration of different wave-bands in a survey. Hidden selection effects may lead to a systematic error, which is not included into eq. 1, and which may simulate large scale inhomogeneities. It is a difficult problem to catch all possible selection effects to be sure in detection of true large scale structures. One way to study an influence of selections is to perform numerical simulations of the procedure of photometric redshift determination taking into account each known selection effect within a specific survey.

Another way to overcome selection effects is to perform many deep multi-band surveys in different directions on the sky and using different instruments and independent methods of photometric redshift estimations. In section below we suggest such observational program for testing the reality of the detected structures.

## 3. General parameters of studied deep fields

Here we analyze and compare redshift distributions in two deep fields: the Hubble Ultra Deep Field (HUDF) [2, 5] and FORS Deep Field (FDF) of the ESO VLT[10, 1].

In Table 2 we present main parameters of deep galaxy surveys, which are used in our analysis of radial galaxy distributions. The transverse sizes which correspond to angular extent of studied deep fields are presented in Tab.3.

| Name | $\alpha$ | $\delta$ | ang.size | $m_{lim}$ |
|---|---|---|---|---|
| HUDF | $03^h32^m$ | $-27°47'$ | $3' \times 3'$ | 29 |
| FDF | $01^h06^m$ | $-25°46'$ | $7' \times 7'$ | 27 |

Table 2. Main parameters of studied deep fields

| z | 1 | 2 | 3 | 4 | 5 |
|---|---|---|---|---|---|
| $l_{HUDF}$ (Mpc) | 2.8 | 4.4 | 5.4 | 6.1 | 6.6 |
| $l_{FDF}$ (Mpc) | 7.5 | 12 | 14 | 16 | 18 |

Table 3. The transverse sizes which correspond to angular extent of studied deep fields.

## 4. Method of detection of super-large structures in galaxy distribution

Our method of detection possible super-large structures is based on analysis of the radial galaxy distribution in several deep galaxy surveys for different directions on the sky. For each field on the sky we perform :

- construction the observed redshift distribution $\Delta N_{obs}(z)/\Delta z$ for several redshift bins $\Delta z$
- construction the redshift distribution $dN_{ml}(z)/dz$ for magnitude limited homogeneous distribution of galaxies in considered deep field
- estimation the expected number fluctuations $\Delta N$ in fixed redshift bins $\Delta z$ for the case of Poisson distribution and the radial distribution of galaxies having power-law two-point correlation function
- extraction the regions of scales where observed density/number fluctuations exceed expected Poisson's $3\sigma$ level for the number fluctuation in a considered redshift bin of the radial distribution of galaxies comparison of radial redshift distribution $dN_{\alpha,~\delta}(z)/dz$ for different directions $(\alpha, \delta)$ on the sky.

Deep galaxy surveys probe very large distance scales and contain large numbers of galaxies even for small angular region on the sky. Hence it is natural to expect the universal galaxy distribution along z-coordinate, which corresponds to a magnitude limited sample observed in a deep galaxy survey.

Redshift distribution for observed magnitude limited galaxy samples is usually approximated by simple empirical formula (see [11, 12, 14]):

$$dN_{ml}(z) = A z^\alpha \exp(\frac{z}{z_0})^\beta dz , \qquad (3)$$

where $\alpha, \beta, z_0$ are free parameters, which should be found by least-squares method, and $A$ is the normalization constant which corresponds the condition $\int dN_{ml} = N_{total-ml}$.

We also have tested how the above eq.(3) represents the magnitude limited samples of mock catalogs, which modeled the geometry and total number of galaxies in the corresponding deep field surveys. We used homogeneous spatial distribution of galaxies with Schechter luminosity function and introduce observed magnitude limit. We used the inverse relation $z = l(z)^{-1}$ to get redshift distribution from radial metric distances. Our numerical results also confirm the validity of the approximation given by eq.(3) for wide class of spatial distributions of galaxies.

The shot Poisson's noise has dispersion:

$$\sigma_p^2 = \frac{\langle N^2 \rangle - \langle N \rangle^2}{\langle N \rangle^2} = \frac{1}{N} , \qquad (4)$$

where the average number of galaxies $\langle N \rangle$ may be calculated by using eq.(3), so that for each bin $dz$ we have $\langle N \rangle = \langle N(dz) \rangle = \langle dN(dz) \rangle = dN_{ml}$. Due to sufficiently large observed number of galaxies in redshift bins $N_{obs}(dz) \approx 100$, for considered surveys the shot noise contribution to observed fluctuations is restricted by small value $\sigma_P \approx 0.1$.

Therefore the main contribution to the observed deflection of galaxy number from expected homogeneous magnitude limited value, given by eq.(3), is determined by the large-scale correlated structures in galaxy spatial distribution.

The expected value of the "cosmic variance" $\sigma_\xi^2$ caused by structures characterized by 2-point correlation function $\xi(r)$, is given by the relation (Peebles 1980; Gabrielli et al. 2005)

$$\sigma_\xi^2 = \frac{1}{V^2} \int_V dV_1 \int dV_2 \xi(|\vec{r}_1 - \vec{r}_2|) \tag{5}$$

where $V$ is the volume of integration, and the argument of the correlation function is $r = |\vec{r}_1 - \vec{r}_2|$.

For spherical volume V and power law 2-point correlation function $\xi(r) = (r_0/r)^\gamma$ there is simple formula, given by Peebles (1980) for expected dispersion $\sigma_\xi^2$, which in our case should be divided by the factor (1+z) to take into account the linear growth of structures. Therefore theoretically expected value of the dispersion may be estimated as:

$$\sigma_{theor}^2 = \frac{J_2}{1+z}\left(\frac{r_0}{r}\right)^\gamma, \tag{6}$$

where constant is given by

$$J_2 = \frac{72.0}{(3-\gamma)(4-\gamma)(6-\gamma)2^\gamma} \tag{7}$$

For the standard value $\gamma = 1.8$, the constant is $J_2 = 1.865$.

For estimation the expected dispersion in number fluctuations within elongated volumes of a deep narrow angle survey we used the effective radius of the studied volume of each redshift bin $dz$ in the form:

$$r = r_{eff} = \left(\frac{3}{4\pi} r^2 \Delta r \Omega\right)^{\frac{1}{3}}, \tag{8}$$

where $\Delta r$ corresponds to the size of the redshift bin $dz$, $\Omega$ is solid angle of the survey. The standard value of the characteristic scale $r_0 = 5 Mpc/h$.

## 5. Searching super-large structures

Let us consider photometric redshift distributions in three galaxy surveys: HUDF and FDF. We compare observed distributions for redshift bins $\Delta z = 0.1, 0.2, 0.3, 0.5$ with expected galaxy number distributions in the case of homogeneous artificial samples which fill the volume of corresponding survey.

As a measure of the deflection of observed galaxy numbers $N_{obs}(z, \Delta z)$ at redshift z within redshift bin $\Delta z$ from homogeneous distribution we shall use the quantity:

$$\sigma_{obs}(z, \Delta z) = \frac{N_{obs}(z, \Delta) - \langle N \rangle}{\langle N \rangle}, \tag{9}$$

where the expected average galaxy number is given by eq.(3), so that $\langle N \rangle = dN_{ml}(z, \Delta z)$.

We also calculate the expected number fluctuations $\sigma_P$ (eq.(4)) and $\sigma_{theor}$ (eq.(6)) for each $(z, \Delta z)$ taking into account the transverse size of a deep field and effective radius of a considered bin, which gives an upper limit for the dispersion (see Somerville et al. 2004). In the Table 4 we present the values of theoretically expected dispersions:

| | HUDF | | | | | | | | |
|---|---|---|---|---|---|---|---|---|---|
| | $\Delta z = 0.2$ | | | $\Delta z = 0.3$ | | | $\Delta z = 0.5$ | | |
| Z | $r_{eff}$ | $\sigma_{theor}$ | $\sigma_P$ | $r_{eff}$ | $\sigma_{theor}$ | $\sigma_P$ | $r_{eff}$ | $\sigma_{theor}$ | $\sigma_P$ |
| 1 | 6.68 | 0.56 | 0.004 | 7.65 | 0.24 | 0.003 | 8.16 | 0.24 | 0.002 |
| 2 | 7.72 | 0.39 | 0.009 | 8.91 | 0.12 | 0.006 | 10.47 | 0.09 | 0.004 |
| 3 | 7.88 | 0.34 | 0.027 | 9.02 | 0.09 | 0.017 | 10.68 | 0.066 | 0.011 |
| 4 | 7.67 | 0.31 | 0.078 | 8.76 | 0.074 | 0.05 | 10.53 | 0.055 | 0.031 |
| 5 | 7.40 | 0.29 | 0.21 | 8.49 | 0.067 | 0.12 | 10.08 | 0.049 | 0.087 |
| | FDF | | | | | | | | |
| | $\Delta z = 0.2$ | | | $\Delta z = 0.3$ | | | $\Delta z = 0.5$ | | |

| Z | $r_{eff}$ | $\sigma_{theor}$ | $\sigma_P$ | $r_{eff}$ | $\sigma_{theor}$ | $\sigma_P$ | $r_{eff}$ | $\sigma_{theor}$ | $\sigma_P$ |
|---|---|---|---|---|---|---|---|---|---|
| 1 | 12.81 | 0.31 | 0.004 | 14.7 | 0.07 | 0.003 | 15.71 | 0.07 | 0.002 |
| 2 | 15.03 | 0.22 | 0.009 | 17.13 | 0.04 | 0.006 | 20.12 | 0.03 | 0.004 |
| 3 | 15.12 | 0.19 | 0.027 | 17.33 | 0.03 | 0.017 | 20.55 | 0.02 | 0.011 |
| 4 | 14.73 | 0.17 | 0.078 | 16.86 | 0.02 | 0.05 | 20.09 | 0.017 | 0.031 |
| 5 | 14.22 | 0.16 | 0.21 | 16.35 | 0.02 | 0.12 | 19.41 | 0.015 | 0.087 |

Table 4. The expected values of the dispersion for different redshift bins at diffrent redshifts for HUDF and FDF deep fields.

Then we extract redshift intervals which correspond over- and under-density regions having number galaxy deflection exceeding $3\sigma$ threshold of Poisson's expected value, so which present the fluctuations due to correlated structures.

We shall call these regions super-large clusters (SLCi) or super-large voids (SLVi). Note that in our definition the SLC is an overdensity region with number density contrast $\Delta N / N > \sigma_P$ and the SLV is not an empty but is an underdensity region with number density $\Delta N / N < -\sigma_P$.

**6. HUDF and FDF surveys**

At figs.5, 6, 7 we present the redshift distributions $dN(z, \Delta z)$ of HUDF and FDF surveys for bins $\Delta z = 0.3, 0.5$. Expected empirical distribution according to eq.(3) is also shown as continuous line.

Observed deviations from expected homogeneous magnitude limited galaxy distribution is presented at figs.1, 2, 3, for the redshift bins $\Delta z = 0.1, 0.2, 0.3$.

*HUDF sample*

The photo-z and redshift distributions in HUDF were considered in details by Coe et al. (2006), where they derived photometric z for 7000 galaxies, by using soft criteria for galaxies extraction.

To get reliable results in photometric redshift distribution here we use our own catalog of photo-z, which is based on only high signal to noise ratio and high quality of photo-z determination. The details of the procedure and the catalog is presented elsewhere (Nabokov & Baryshev 2008). Our catalog contain 2700 galaxies.

Intriguingly for all redshift bins the redshift intervals of overdensity and underdensity regions are stable and demonstrate coherent behavior at super-large scales. We identify the following regions:

| Name | z interval | Size (Mpc) |
|---|---|---|
| HUDF-SLC-1 | 0÷1.2 | 3724 |
| HUDF-SLV-1 | 1.2÷2.2 | 1647 |
| HUDF-SLC-2 | 2.2÷3.1 | 939 |
| HUDF-SLV-2 | 3.1÷4.2 | 796 |
| HUDF-SLC-3 | 4.2÷4.7 | 330 |
| HUDF-SLV-3 | 4.7÷6 | 586 |

Table 5. HUDF super-large scales for bin size $\Delta z = 0.3$

*FDF sample*

About 7000 photo-z for FDF survey was presented in Appenzeller et al.(2004), where redshift distribution was constructed for $\Delta z = 0.1$. We have used these data to construct redshift distributions and observed deviations for the FDF sample with redshift bins $\Delta z = 0.2, 0.3, 0.5$.

As in the HUDF case the FDF redshift distributions demonstrate coherent behaviour at super-large scales for the redshift intervals of overdensity and underdensity regions. We identify the following regions:

| Name | z interval | Size (Mpc) |
|---|---|---|
| FDF-SLC-1 | 0÷0.83 | 2791 |
| FDF-SLV-1 | 0.83÷2.12 | 2412 |
| FDF-SLC-2 | 2.12÷3.75 | 1797 |
| FDF-SLV-2 | 3.75÷4.18 | 293 |
| FDF-SLC-3 | 4.18÷5.15 | 546 |
| FDF-SLV-3 | 5.15÷6 | 383 |

Table 5. FDF super-large scales for bin size $\Delta z = 0.3$

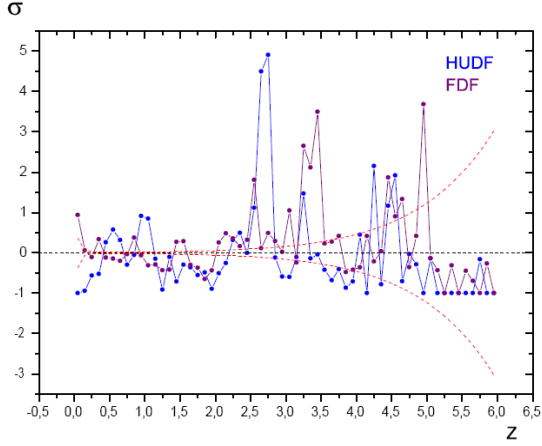

Fig.1. Observed deviations $\sigma_{obs}(z, \Delta z)$ in case $\Delta z = 0.1$ for HUDF and FDF data. Dashed lines correspond to Poisson's expected deviations at $3\sigma$ level.

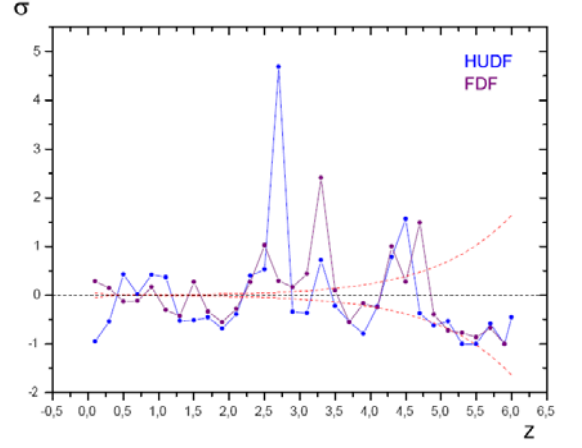

Fig.2. Observed deviations $\sigma_{obs}(z, \Delta z)$ in case $\Delta z = 0.2$ for HUDF and FDF data. Dashed lines correspond to Poisson's expected deviations at $3\sigma$ level.

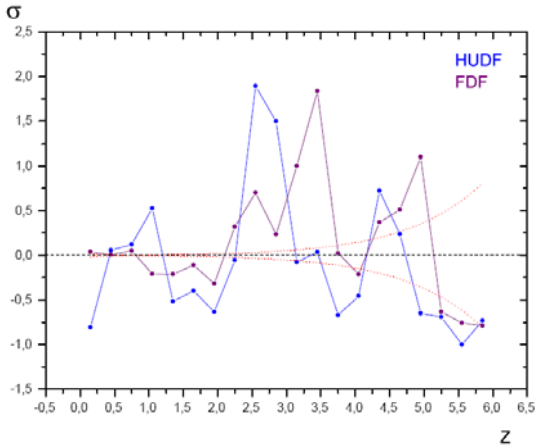

Fig.3. Observed deviations $\sigma_{obs}(z, \Delta z)$ in case $\Delta z = 0.3$ for HUDF and FDF data. Dashed lines correspond to Poisson's expected deviations at $3\sigma$ level.

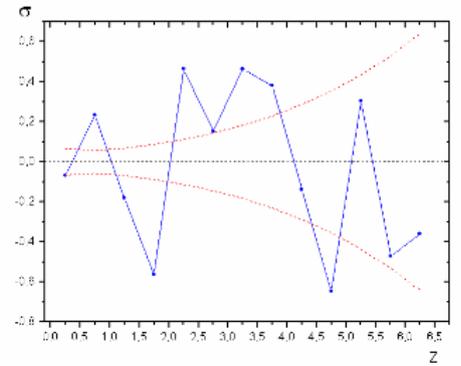

Fig.4. Observed deviations $\sigma_{obs}(z, \Delta z)$ in case $\Delta z = 0.5$ for GRB data. Dashed lines correspond to Poisson's expected deviations at $\sigma$ level.

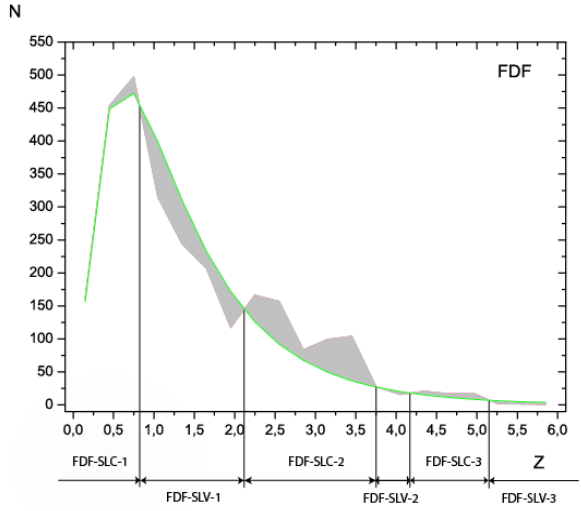

Fig.5. Observed redshift distribution in the case $\Delta z = 0.3$ for FDF data. Continuous line corresponds the expected redshift distribution in homogeneous magnitude limited sample.

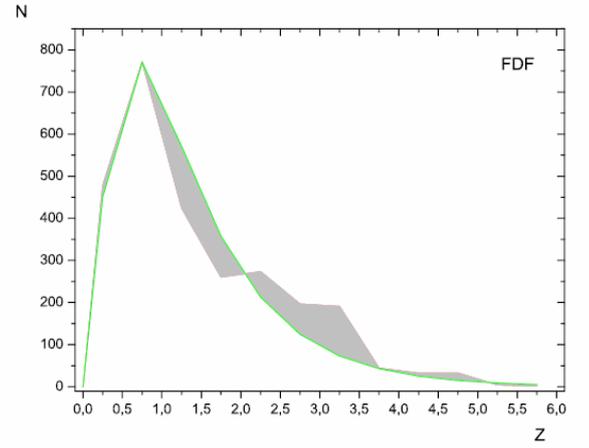

Fig.6. . Observed redshift distribution in the case $\Delta z = 0.5$ for FDF data. Continuous line corresponds the expected redshift distribution in homogeneous magnitude limited sample.

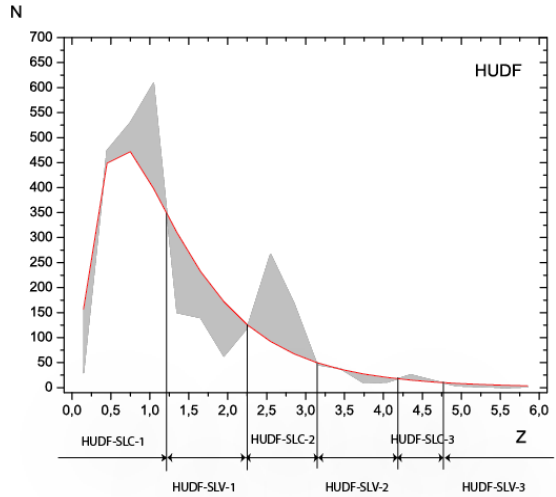

Fig.7. Observed redshift distribution in the case $\Delta z = 0.3$ for HUDF data. Continuous line corresponds the expected redshift distribution in homogeneous magnitude limited sample.

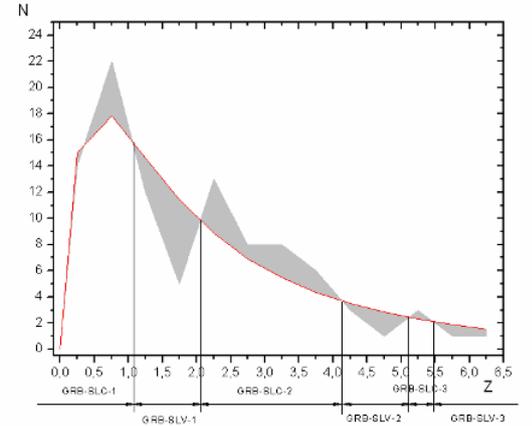

Fig.8. Observed redshift distribution in the case $\Delta z = 0.5$ for GRB data. Continuous line corresponds the expected redshift distribution in homogeneous magnitude limited sample.

*GRB sources*

Gamma ray bursts of long-type occur in very distant galaxies during the supernova explosion. Observations of spectral lines in afterglows and in the host galaxies give a sample of very distant galaxies, where SN explosions were detected. In the paper by Coward et al.(2008), "Where are the missing gamma ray burst redshifts?", was found, that in the redshift distribution of 69 spectroscopically measured redshifts obtained following Swift localizations, there is a strong minimum at $z \approx 2$ .

In figs.4, 8 we present the redshift distribution and the deviations from homogeneous sample for 100 spectroscopically measured redshifts available at http://heasarc.gstc.nasa.gov/docs/swift . Intriguingly the same minimum is seen in the photo-z distribution of galaxies, which we identify with super-large underdense regions HUDF-SLV-1 and FDF-SLV-1. Hence in our interpretation the missing gamma ray redshifts are absent simply due to the deficit of galaxies at this redshift interval. Note that other two peaks at redshift 2.5 and 3.5 seen in GRB redshift distribution also exist in our samples of HUDF and FDF galaxy distributions.

## 7. Covering sky in different directions

The angle on the sky between the HUDF and the FDF directions is about 36 degrees, hence the transverse size of a super-large structure which is seen in both fields at redshift 1 is about 1600 Mpc/h.

From redshift distributions and number deviations presented for HUDF and FDF surveys we conclude, that there are the same super-large structures which intersect line off-site of considered fields and which have slightly different amplitudes and relative shifts in redshifts. The sizes of detected super-large structures in radial direction (about 1000 Mpc) are consistent with the transverse sizes.

Observational test on the reality of the supper-large structures may be obtained by organizing sky covering net (sells about 10n x 10n degrees) of very deep narrow angle (1n x 1n arc-minutes) multi-band photometric surveys of galaxies, which are achievable for large ground based telescopes. As a result of such observational program one can construct the 3-dimensional map of neighboring "metagalaxies". Another test may be performed when the number of GRB redshift will be sufficiently large for the study of redshift distributions in different directions at the sky.

## 8. Conclusions

In this paper we presented analysis of the redshift distribution of galaxies from very deep surveys (HUDF, FDF). Number galaxy fluctuations in large redshift bins essentially exceed the $3\sigma$ level and hence may be caused by correlated structures. Though possibility of hidden systematic effects is still needed to be further studied, here we presented observational evidences on the reality of super-large structures with sizes about 1000 Mpc/h in the luminous matter distribution in the Universe.

Our result is in agreement with recent discoveries of real large structures in the Universe, obtained by different observational approaches. For example, there is known structure having size about 500 Mpc discovered in SDSS (Sloan Great Wall - Gott et al. 2005). Sylos Labini et al. (1998) noted possible structures with sizes of 1000 Mpc by using analysis of large redshift surveys. Padmanabhan et al. (2007) discovered excess of power spectrum of SDSS luminous red galaxies using photometric redshifts, which indicate presence of structures with wave number k=0.005 corresponding linear size λ=1200 Mpc. Miller et al. (2004) discovered 200 - 300 Mpc size super clusters in 2dF QSO samples. Brand et al. (2003) found 100-Mpc-scale structures in three-dimensional distribution of radio galaxies. Rudnick et al. (2007) detected a dip in number counts of NVSS sources and corresponding the WMAP cold spot, which indicates a presence of a large void with size 300 Mpc. Redshift distribution of GRB host galaxies consistent with redshift distributions in HUDF and FDF deep fields.

To further study of the reality of the super-large structures we suggested an observational test by constructing photometric redshift distributions in narrow very deep multi-color galaxy surveys, which cover the sky by 10x10 degrees net. In the beginning of the 20th century, using the largest at that time telescopes Edwin Hubble opened the door into the "realm of galaxies", and now, in the beginning of the 21st century, by operating with modern largest telescopes we have opportunity to observational study the "realm of metagalaxies".